\documentclass[prl,aps,twocolumn,floats,superscriptaddress]{revtex4}
\usepackage{graphicx}

\newcommand{\be}{\begin{equation}}
\newcommand{\ee}{\end{equation}}
\newcommand{\bea}{\begin{eqnarray}}
\newcommand{\eea}{\end{eqnarray}}

\begin{document}

\preprint{APS/123-QED}

\title{Disappearance of the metal-like behavior in GaAs two-dimensional holes below 30\,mK}

\author{Jian Huang}

\affiliation{%
PRISM, Princeton University, Princeton, NJ 08544, USA\\}%
\author{J. S. Xia}
\affiliation{%
Department of Physics, University of Florida, Gainesville, FL USA\\}

\author{D. C. Tsui}
\affiliation{%
Department of Electrical Engineering, Princeton University,
Princeton, NJ 08544, USA\\}

\author{L. N. Pfeiffer}%
\author{K. W. West}%
\affiliation{%
Bell Labs, Lucent Technologies, Murray Hill, NJ 07974, USA\\}
\date{\today}
\begin{abstract}

The zero-field temperature-dependence of the resistivity of
two-dimensional holes are observed to exhibit two qualitatively
different characteristics for a fixed carrier density for which only
the metallic behavior of the so-called metal-insulator transition is
anticipated. As $T$ is lowered from 150\,mK to 0.5\,mK, the sign of
the derivative of the resistivity with respect to $T$ changes from
being positive to negative when the temperature is lowered below
$\sim$30\,mK and the resistivity continuously rises with cooling
down to 0.5\,mK, suggesting a crossover from being metal-like to
insulator-like.

\end{abstract}

\pacs{Valid PACS}
\maketitle

Charges in an electronic system are subjected to the influences of
the disordered environment and interaction among themselves. Early
study on the effects of disorder, ignoring interaction, revealed
that a conducting system undergoes a qualitative change into an
insulator when disorder is sufficiently large~\cite{Anderson'58}. On
the other hand, the effects due to interaction, e.g. Wigner
Crystallization, turn out to be more complicated than
expected~\cite{AAL,Fin,Cas,Stern-Das,gold,Das,zna}. The interplay of
disorder and interaction, one of the most important and fascinating
subjects in condensed matter physics, remains to be further
explored.

In two-dimensional (2D) systems, despite the theoretical prediction
of an insulating ground state (even the disorder is
minimal)~\cite{stl,AAL},
experimental discovery of the metal-to-insulator transition
(MIT)~\cite{mit1} at finite temperatures ($T$) has stimulated
further interest of investigating the interaction effects as the
primary cause of MIT. Theoretical models include scattering of
charges by the potential created by all other charges and
impurities, and, recently, scaling theory in a multi-valley
system~\cite{Punn-Finkel}.
If there is a true metallic ground state, it would be a strong
indication that there exists a Quantum Phase
Transition(QPT)~\cite{qpt,qpt-1,AKS-review,altshuler-review}.
Whether or not the observed finite-$T$ metal-like behavior
necessarily indicates a metallic ground state is thus at the heart
of the question.

Even though the metal-like phenomenon has been demonstrated in
various high quality 2D systems in the $T$-dependence of the
resistivity $\rho$ at temperatures as low as $\sim$30\,mK, there
lacks evidence, due to the experimental limitations, on whether the
metallic signature $d\rho/dT>0$ will survive at even lower
temperatures~\cite{hamilton}. A desired experiment would require the
following two factors at the same time: a 2D sample with very high
quality; and the ability of cooling 2D charges down to mK level.
Performing mK-level $T$-dependence measurements has proven a great
challenge because it demands effective cooling and noise filtering,
and, moreover, a mK-level thermometry for the 2D carriers which is
not yet established. An important progress in techniques of cooling
2D charges is achieved previously by Xia {\it et al}~\cite{xia} who
have demonstrated the cooling of 2D electrons down to 4\,mK. Now, on
the sample side, the GaAs HIGFET adopted in recent
experiments~\cite{noh,jian-1,jian-2,lilly} has shown exceptional
quality due to the fact that there is no intentional doping in the
system, thus a record-low concentration of $6\times 10^{8}$
cm$^{-2}$ has been achieved. Novel non-activated
behaviors~\cite{jian-1} and direct evidence of interaction
effects~\cite{jian-2} have been recently reported by using such
HIGFET devices.

The present work combines both the exceptionally-clean HIGFET device
with the unique cooling techniques. In a similar nuclear
demagnetization refrigerator as that in Ref.~\cite{xia}, we have
performed transport measurements on high quality 2D holes at
ultra-low temperatures with a minimum bath $T$ of 0.4\,mK. The
hole-density is varied by a metal gate from $\sim5$ to $8\times
10^{9}$ cm$^{-2}$ for which the usual metal-like behavior is
anticipated. Remarkably, following the metal-like behavior
($d\rho/dT>0$) at $T>$30\,mK, a rising of $\rho$ with cooling is
observed at $T$ below 30\,mK, signifying a crossover into an
insulator-like regime ($d\rho/dT<0$). The increase of the
resistivity with cooling persists down to the lowest bath $T$ of
0.5\,mK.

\begin{figure}[t]
\includegraphics[width=2.55in,trim=0.07in 0.1in 0.02in 0.0in]{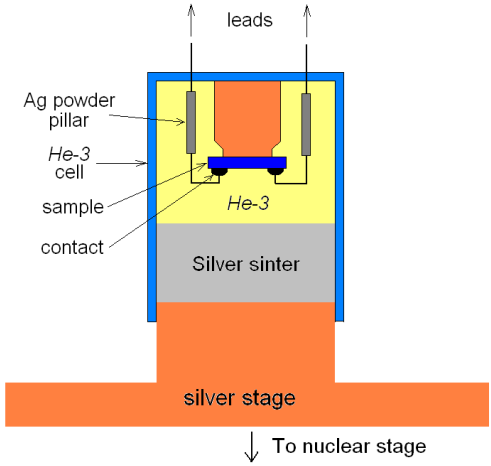}
\caption{\label{fig:demag} (color online) Schematics of the cooling
arrangements. The sample sits in a liquid $^3$He cell (located on a
silver base) which is connect to the mixing chamber through a copper
band. The wires are connected to the sample contacts through some
silver power pillars that are inside the cell.}
\end{figure}

The sample preparation process and measurement details can be found
in ref.~\cite{jian-1}. Fig.~\ref{fig:demag} is an illustration of
the cooling strategies inside the fridge. The sample sits inside a
cell, filled with liquid $^3$He, mounted on a silver base which is
connected to the nuclear stage. Inside the cell, electrodes on the
sample are first connected to short gold wires through silver paste
(black dots on the sample). The gold wires are then attached to the
sintered silver powder pillars each of which has one M$^2$ surface
area. These pillars not only provide much more efficient heat
exchange with the bath, but also serve as filters against the
microwave noises. $T$ is measured by a $^3$He melting curve
thermometer (MCT) located on the same silver base as the $^3$He
sample cell (not shown).

\begin{figure}[b]
\includegraphics[width=3.1in,trim=0.07in 0.0in 0.02in 0.3in]{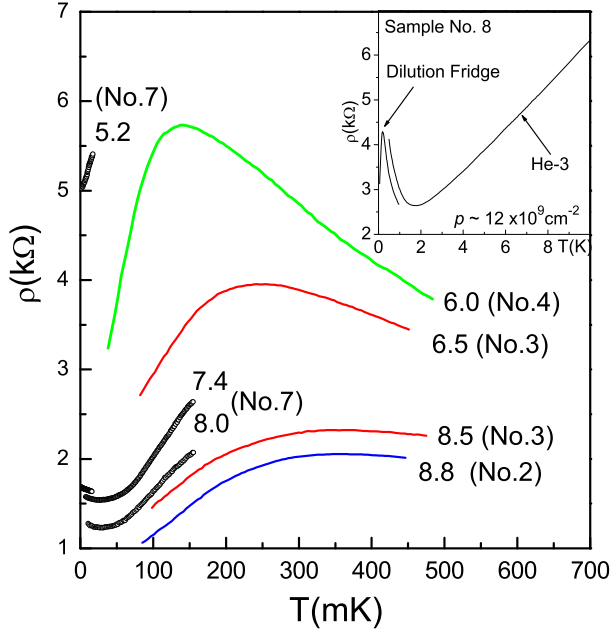}
\caption{\label{fig:rt-0} (color online) Zero-field $T$-dependence
of $\rho(T)$ from five different samples (No.2, 3, 4, 7, and 8) for
densities from $p=5.2$ to $8.8\times10^{9}$ $cm^{-2}$. Inset:
$\rho(T)$ from sample No.8 for $p=10\times10^{9}$ $cm^{-2}$ at
$0.03$\,K$\leq T\leq10$\,K.}
\end{figure}

\begin{figure}[bh]
\includegraphics[width=3.15in,trim=0.07in 0.0in 0.02in 0.22in]{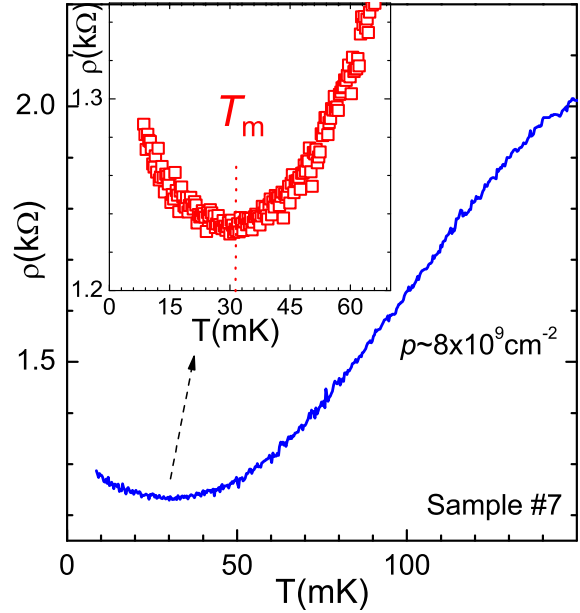}
\caption{\label{fig:rt-1} (color online) Zero-field $T$-dependence
$\rho(T)$ for $p=8\times10^{9}$ $cm^{-2}$ ($r_s\sim33$) at a
temperature range from 150\,mK to 8\,mK. Inset: Zoom-in of the
$\rho(T)$ at low $T$ with a minimum around 32\,mK (dotted line).}
\end{figure}

\begin{figure}[b]
\includegraphics[width=3in,trim=0.07in 0.0in 0.02in 0.15in]{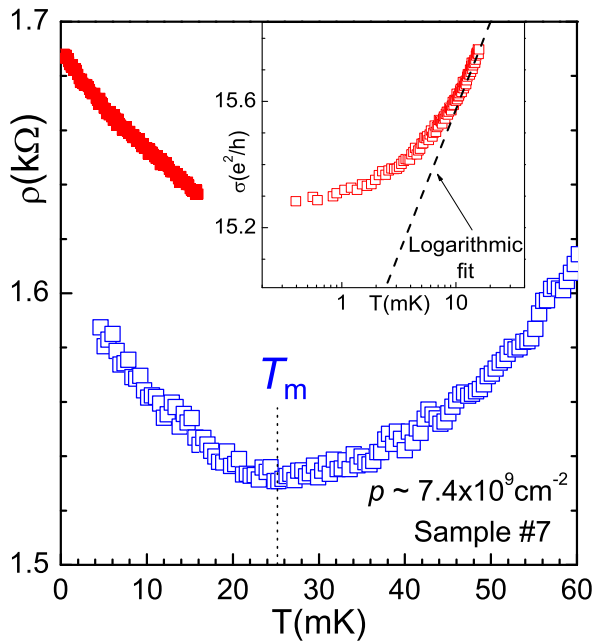}
\caption{\label{fig:rt-2} (color online) $T$-dependence of $\rho(T)$
for $p=7.4\times10^{9}$ $cm^{-2}$ from 0.5 \,mK to 60\,mK. Data
before demagnetization (scattered points) corresponds to a slight
different density than that after demagnetization due to the turning
on/off the gate voltage. $T_m\sim$25\,mK(dotted line). Inset:
conductivity ($\sigma=1/\rho$) plotted versus $\ln T$ with
logarithmic fit (solid line) performed for 10\,mK $\lesssim
T\lesssim20$\,mK.}
\end{figure}
\begin{figure}[t]
\includegraphics[height=3.2in,trim=0.07in 0.05in 0.12in 0.2in]{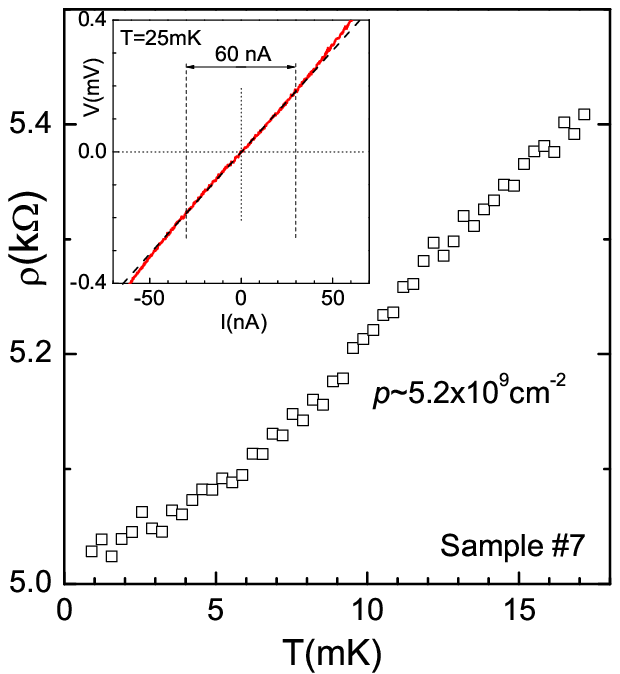}
\caption{\label{fig:rt-3} (color online) $T$-dependence of the
resistivity ($\rho(T)$) for $p=8\times 10^{9}$ cm$^{-2}$. Inset: the
dc-IV result at 25\,mK showing a linear response window from -30\,nA
to 30\,nA.}
\end{figure}
Before getting into details of the lowest-$T$ data, a comparison is
first drawn with the previous results (Ref.~\cite{jian-1,jian-2})
obtained at $T\geq30$\,mK. Fig.~\ref{fig:rt-0} shows the
$T$-dependence of the resistivity $\rho(T)$ for various 2D-hole
densities ($p=5.2$ to $12\times10^{9}$ $cm^{-2}$) from five
different samples, including sample No.7 which will be the focus.
Starting from the inset, $\rho(T)$ from sample No.8 is shown for a
temperature range from 0.04 to 10\,K. The rising of $\rho(T)$ with
increasing $T$ at $T>1.5$\,K (estimated Fermi temperature
$T_F\sim$~1.5\,K) is due to the dominating electron-phonon
interaction so the system is classical. As $T$ is lowered below
$1.5$\,K, the system undergoes a classical-quantum crossover into
the quantum regime where $\rho$ first rises with cooling in an
approximate logarithmic fashion. This behavior is confirmed in the
dilution-fridge data (for a slightly different density) in the
overlapped temperature range (0.5\,K to 1\,K). As for lower $T$,
$\rho(T)$ exhibits the characteristic metal-like downward bending
starting from the peak around 250\,mK. Now, Fig.~\ref{fig:rt-0}
provides a more detailed view of $\rho(T)$ at temperatures below
500\,mK for various labeled densities measured in different samples.
For samples No.2, 3, and 4, as $p$ is lowered, the metallic bending
gets stronger and the resistivity peak moves towards lower $T$. For
sample No.7, $\rho(T)$ for $p=7.4$ and $8.0\times10^{9}$ $cm^{-2}$
agrees well with these results in the overlapped temperature region
(up to 160\,mK). However, the lower-$T$ results, which we discuss
for the rest of the paper, are qualitatively different from being
metal-like.

Fig.~\ref{fig:rt-1} shows the $T$-dependence of $\rho(T)$ of sample
No.7 from 8\,mK to 150\,mK for a 2D-hole density $p=8.0\times
10^{9}$ cm$^{-2}$. At $T>65$\,mK, $\rho(T)$ shows the usual
metal-like ($d\rho/dT>0$) behavior. However, when $T$ is lowered
below 65\,mK, $\rho(T)$ shows weaker $T$-dependence with cooling and
reaches a minimum ($d\rho/dT=0$). As $T$ is further reduced,
$\rho(T)$ starts to rise, faster with cooling, all the way down to
8\,mK, so that the sign of $d\rho/dT$ becomes negative. The inset is
a blow-up of the low-$T$ $\rho(T)$ showing the minimum at
$T_{m}=$32\,mK. $\rho(T)$ rises from $\sim1.22$\,k$\Omega$ to
$\sim1.28$\,k$\Omega$ from 32\,mK to 9\,mK. The measurements were
repeated several times with different set of leads and different
current drives ($I_{drive}$) from 0.5\,nA to 1\,nA, which correspond
to a change of power from 0.25 and $1\times 10^{-15}W$.
Notice that the linear response window for $p=5.2\times 10^{9}$
cm$^{-2}$, shown later in the inset of fig.~\ref{fig:rt-3}, is $\pm
30nA$. For $p=8.0\times 10^{9}$ cm$^{-2}$, this window is much wider
so our current drives, $\leq$1\,nA, are by far smaller. The results
remain consistent with variation of the drives, indicating the Joule
heating is negligible.

Fig.~\ref{fig:rt-2} is the $T$-dependence data for $p=7.4\times
10^{9}$ cm$^{-2}$ with the $T$-range set from $0.5$\,mK to 60\,mK
for a closer look. The scattered data points (from 8\,mK to 60\,mK)
are collected before demagnetization. Solid line (from 0.4\,mK to
16\,mK) is obtained after demagnetization for a slightly different
density. $\rho(T)$ exhibits similar behavior as that for $p=8\times
10^{9}$ cm$^{-2}$ except that the minimum appears at a lower $T_{m}$
of 25\,mK. Now, at $T\leq$\,25mK, $\rho(T)$ rises with cooling
consistently down to the minimum bath temperature of $\sim0.5$\,mK.
The increase rate of $\rho$ versus decreasing $T$
($\sim60\Omega/16$\,mK), slightly larger at $T\lesssim$6\,mK, is
very similar to that for $p=8.0\times 10^{9}$ cm$^{-2}$ shown in
Fig.~\ref{fig:rt-0}. The inset shows the conductivity-$\ln T$
relation which is different than the logarithmic corrections
predicted for both the weak-localization and the weak interaction
scenarios. The dash line, fitted for $5$\,mK$\leq T\leq 15$\,mK,
follows the logarithmic function: \be \label{log} {\sigma(T)}
=\sigma_0+\frac{Ce^2}{\pi^2\hbar}ln(T/T_0) \,\ee with
$\sigma_0\sim10e^2/h$, $T_0\sim10\time10^{-6}$\,mK, and $C\sim4.2$
which are quite different than previous findings~\cite{Pud,Simmons}.
Here, we stress on two factors: First, the interaction is not weak.
With $r_s$ approximately $35$ (assuming an effective mass of
$0.35m_0$), the interaction-induced modification to $\rho(T)$ could
be qualitatively different than perturbations~\cite{zna,AKS-review}.
The second factor, perhaps more important from the experimental
point of view, is related to the possible heating effects (discussed
later).

In fig.~\ref{fig:rt-3}, the $T$-dependence is shown for $p=5.2\times
10^{9}$ cm$^{-2}$ which already approaches the critical density
($\sim4\times 10^{9}$ cm$^{-2}$) of MIT found in such devices.
comparing with the results for $p=8.0$ and $7.4\times 10^{9}$
cm$^{-2}$, $\rho(T)$ for this density remains decreasing with $T$
consistently down to 5\,mK, and then shows weaker variation at lower
$T$ and tends to saturate around the base temperature of 0.5\,mK. No
upturning occurred. The rate at which $\rho(T)$ decreases,
$\sim0.3k\Omega/10$\,mK, is approximately 3 times faster than that
for $p=8.0\times 10^{9}$ cm$^{-2}$. The dc I-V result obtained at
25\,mK is shown as the inset with the slope corresponding to a
resistivity of $\sim5.7$\,k$\Omega$.

Should the increase of $\rho(T)$ with cooling observed for $p=8.0$
and $7.4\times 10^{9}$ cm$^{-2}$ indicate a possible insulating
ground state? In order to rigorously establish an insulator, $\rho$
should grow towards infinity as $T$ approaches zero. Though our
results show the right sign of an insulator ($d\rho/dT<0$),
$\rho(T)$ tends to approach finite values towards $T=0$. This is
very likely due to the heating effects that cause $T_{2D}$($T$ of 2D
holes) lag $T_{bath}$, and the difference between the two may grow
as $T_{bath}$ approaches the base.
Since the phonon modes are severely suppressed, the heat generated
in the 2D system is carried out primarily through charge diffusion
across the contacts. The cooling of the contact pads is limited by
the Kapitza resistance $\sim T^{-1}$ (between the $^3$He atoms and
the gold contacts and wires) which becomes significant at $\sim$\,mK
level, producing temperature discontinuity across the boundaries.
Another possible factor is there could be other processes that can
exchange energy with the 2D holes. For example, the hyperfine
interactions can limit the temperature since the interaction energy
$g^*\mu_B B_{nu}\sim 0.1$\,mK to $\sim1.4$\,mK ($\mu_B$ is the spin
of a carrier and $B_{nu}$ is the effective field due to nuclei).
Also, there are indications of possible spin order occurring as the
charge density approaches the critical density of MIT~\cite{jian-3},
which, in turn, polarize the nuclei spins. All these factors can
interfere with the $T$-dependence measurement and the knowledge of
the actual $T_{2D}$ demands more sophisticated thermometry of the 2D
holes at mK level which is beyond the scope of this paper. Despite
the uncertainty related to the thermometry, the consistent increase
of $\rho$ with cooling at low-$T$ indicates that the 2D holes are
definitely being cooled even when the bath reaches its minimum $T$.


We point out that the increase of the resistance at low-$T$ probably
has a different origin than that resulted from weak-localization
(WL) contribution~\cite{Simmons} observed for much higher densities.
The $T_{m}$ in Ref.~\cite{Simmons} appears at 300\,mK which is one
order higher. Moreover, our results, contrary to
Ref.~\cite{Simmons}, show decrease of $T_{m}$ with reducing $p$,
which possibly indicates a different quantum effect. Also, the
values of $\rho$ in our case is much smaller: i.e. for $p=8.0\times
10^{9}$ cm$^{-2}$, $\rho\sim1.2k\Omega\sim0.05h/e^2$ which is
$\sim6$ times smaller than that for a 6 times higher $p$ in
Ref.~\cite{Simmons}. This transition from being metal-like to being
insulating at $T\lesssim25$\,mK can not be explained by the existing
Fermi-Liquid-based models~\cite{Fin,Stern-Das,zna,Das} (e.g. for a
fixed density, the $T$-dependence of scattering by charges and
impurities only leads to monotonically reduced resistivity with
cooling in the metal-like regime($d\rho/dT>0$)~\cite{Das,zna}).
Notice that the estimated $T_F$ ($\sim 600$\,mK) is significantly
higher than our temperature range. Also, the scaling theory
including the valleys of Ref.~\cite{Punn-Finkel} leads to, in the
diffusive limit($k_BT\ll\hbar/\tau$, $\tau\sim$ diffusing time), a
quantum critical point (QCP) that separates the two phases, and the
crossing from one phase to the other requires change of interaction
for fixed disorder. Here, our results, though also into the
diffusive regime since $T\ll\hbar/\tau\sim100$\,mK, are from a
single-valley-system.

For $p=5.2\times 10^{9}$ cm$^{-2}$, the steeper metal-like decrease
of $\rho(T)$ moves towards lower $T$, which seems in agreement with
the statement made in Ref.~\cite{Das} due to the reduced $E_F$.
Since the electron-electron interaction is now much stronger
($r_s\sim50$), the many-body coherence length $l_{\phi}$ is much
reduced so that $T_{m}$ could be lower than $T_{2D}$. The slight
$T$-dependence for $T_{bath}\leq3$\,mK probably indicates that
$\rho(T)$ eventually rises at lower temperatures when $l_{\phi}$
becomes sufficiently large, consistent with earlier results.

To summarize, for a fixed carrier density around $8\times 10^{9}$
cm$^{-2}$, we have observed a crossover from the metal-like state
($d\rho/dT>0$) into an insulator-like state when $T$ is lowered
below 30\,mK, and this insulator-like behavior($d\rho/dT<0$)
prevails down to a base temperature of 0.5\,mK.
These results suggest that the metal-like behavior of MIT is likely
to be a finite-temperature effect and the ground state is possibly
insulating.
\begin{acknowledgments}
We acknowledge the helpful discussions with David Huse and Philip
Anderson. The work at Princeton University is funded by a US NSF
MRSEC grant DMR-0213706. Jian Huang was supported by a DOE grant
DEFG02-98ER45683.
\end{acknowledgments}

\end{document}